\pdfoutput=1

\documentclass[10pt,journal,compsoc]{IEEEtran}

\usepackage[utf8]{inputenc}
\usepackage[T1]{fontenc}
\usepackage[french,english]{babel}

\usepackage{courier}  

\usepackage{listings}  
\usepackage{xcolor}  
\usepackage{textcomp}  

\makeatletter
\let\@ORGmakecaption\@makecaption
\long\def\@makecaption#1#2{\@ORGmakecaption{#1}{#2}\vskip\belowcaptionskip\relax}
\makeatother

\usepackage{flafter}  
\usepackage{balance}  

\usepackage{afterpage}  

\newcommand{\ShiftFloatSecondCol}[1]{\afterpage{#1}}

\usepackage{amsmath}
\usepackage{amssymb}  
\usepackage{dsfont}  
\usepackage{mathabx}  
\usepackage[
separate-uncertainty=true,  
retain-explicit-plus=true  
]{siunitx}

\usepackage{letltxmacro}
\makeatletter
\let\oldr@@t\r@@t
\def\r@@t#1#2{%
	\setbox0=\hbox{$\oldr@@t#1{#2\,}$}\dimen0=\ht0
	\advance\dimen0-0.2\ht0
	\setbox2=\hbox{\vrule height\ht0 depth -\dimen0}%
	{\box0\lower0.4pt\box2}}
\LetLtxMacro{\oldsqrt}{\sqrt}
\renewcommand*{\sqrt}[2][\ ]{\oldsqrt[#1]{#2}}
\makeatother

\ifCLASSOPTIONcompsoc
  \usepackage[nocompress]{cite}
\else
  \usepackage{cite}
\fi
\ifCLASSINFOpdf
   \usepackage[pdftex]{graphicx}
   \graphicspath{{./res/img/}}
   \DeclareGraphicsExtensions{.pdf,.png}
\else
\fi
\ifCLASSOPTIONcompsoc
  \usepackage[caption=false,font=footnotesize,labelfont=sf,textfont=sf]{subfig}
\else
  \usepackage[caption=false,font=footnotesize]{subfig}
\fi
\usepackage{dblfloatfix}
\usepackage{url}
\newcommand\MYhyperrefoptions{bookmarks=true,bookmarksnumbered=true,
pdfpagemode={UseOutlines},plainpages=false,pdfpagelabels=true,
colorlinks=true,linkcolor={black},citecolor={black},urlcolor={black},
}
\ifCLASSINFOpdf
\usepackage[\MYhyperrefoptions,pdftex]{hyperref}
\else
\usepackage[\MYhyperrefoptions,breaklinks=true,dvips]{hyperref}
\usepackage{breakurl}
\fi
\usepackage[all]{hypcap}  

\usepackage{microtype}  

\makeatletter
\long\def\@IEEEtitleabstractindextextbox#1{\parbox{0.922\textwidth}{#1}}
\makeatother

\newcommand\CCTitle{Spherical k-Nearest Neighbors Interpolation}
\newcommand\CCAuthorName{Philippe~Trempe}
\newcommand\CCAuthorTitle{\IEEEmembership{Member,~IEEE,~Jr.~Eng.,~OIQ,~M.~Sc.~A.}}
\newcommand\CCSubject{Geospatial data transformation}
\newcommand\CCKeywords{Geospatial,data,transformation,interpolation,spherical,k,nearest,neighbors,analysis,SkNNI,NISD,NDD,NDDNISD}

\newcommand\CCPelmorexCorp{\mbox{Pelmorex Corp.}}
\newcommand\CCSkNNIlong{Spherical k\nobreakdash-Nearest Neighbors Interpolation}
\newcommand\CCSkNNI{\mbox{SkNNI}}
\newcommand\CCNDDNISDlong{Neighborhood Distribution Debiased Normalized Inverse Squared Distance}
\newcommand\CCNDDNISD{NDDNISD}
\newcommand\CCNISDlong{Normalized Inverse Squared Distance}
\newcommand\CCNISD{NISD}
\newcommand\CCNDDlong{Neighborhood Distribution Debiasing}
\newcommand\CCNDD{NDD}
\newcommand\CCAMERPElong{Absolute Maximum Error Ratio Percentage Error}
\newcommand\CCAMERPE{AMERPE}

\newcommand\CCSknniRepositoryUrl{\texttt{\url{https://ptrempe.page.link/sknni/}}}

\hypersetup{
	pdftitle={\CCTitle},
	pdfsubject={\CCSubject},
	pdfauthor={\CCAuthorName},
	pdfkeywords={\CCKeywords},
	bookmarksnumbered,
	pdfstartview={FitV},
	hidelinks,
	linktoc=all
}

\newcommand\CCFigure[5]{%
\begin{figure}[!t]
	\centering
	\includegraphics[#5]{#2}
	\par\vspace*{-0.75em}
	\caption[#3]{\textbf{#3.} #4}
	\label{#1}
\end{figure}}

\newcommand\CCWideFigure[5]{%
\begin{figure*}[!t]
	\centering
	\includegraphics[#5]{#2}
	\par\vspace*{-0.75em}
	\caption[#3]{\textbf{#3.} #4}
	\label{#1}
\end{figure*}}

\newcommand\CCMultiFigureSub[4]{%
\subfloat[#3]{\includegraphics[#4]{#2}%
\label{#1}}}

\newcommand\CCMultiFigure[4]{%
\begin{figure*}[!t]
\centering
#2
\caption[#3]{\textbf{#3.} #4}
\label{#1}
\end{figure*}}

\newcommand\CCRefsec[1]{\hyperref[#1]{§\thinspace\ref{#1}}}
\newcommand\CCRefSec[1]{\hyperref[#1]{§\thinspace\ref{#1}}}

\newcommand\CCRefexp[1]{\hyperref[#1]{expression~\eqref{#1}}}
\newcommand\CCRefExp[1]{\hyperref[#1]{Expression~\eqref{#1}}}

\newcommand\CCRefeqn[1]{\hyperref[#1]{equation~\eqref{#1}}}
\newcommand\CCRefEqn[1]{\hyperref[#1]{Equation~\eqref{#1}}}

\newcommand\CCReffig[1]{\hyperref[#1]{figure~\ref{#1}}}
\newcommand\CCRefFig[1]{\hyperref[#1]{figure~\ref{#1}}}

\newcommand\CCReftab[1]{\hyperref[#1]{table~\ref{#1}}}
\newcommand\CCRefTab[1]{\hyperref[#1]{Table~\ref{#1}}}

\newcommand\CCReflst[1]{\hyperref[#1]{listing~\ref{#1}}}
\newcommand\CCRefLst[1]{\hyperref[#1]{Listing~\ref{#1}}}

\begin{document}
%
\title{\CCTitle}
%
%
%
%

\author{\CCAuthorName,~\CCAuthorTitle
\IEEEcompsocitemizethanks{\IEEEcompsocthanksitem P. Trempe was with the Department
of Computer and Software Engineering, Polytechnique Montréal (Université de Montréal), Montréal,
Québec, Canada.\protect\\
E-mail: pht@ieee.org
}%
}%

\IEEEtitleabstractindextext{%
\begin{abstract}
Geospatial interpolation is a challenging task due to real world data often being sparse, heterogeneous and inconsistent. For that matter, this work presents \CCSkNNI{}, a spherical interpolation algorithm capable of working with such challenging geospatial data. This work also presents \CCNDDNISD{} an accurate and efficient interpolation function for \CCSkNNI{} which shines due to its spatial awareness in terms of proximity and distribution of observation neighbors. \CCSkNNI{}'s open source implementation is also discussed and illustrated with a simple usage example.
\end{abstract}

}%

\maketitle

\IEEEdisplaynontitleabstractindextext

%
\IEEEpeerreviewmaketitle

\ifCLASSOPTIONcompsoc
\IEEEraisesectionheading{\section{Introduction}\label{sec:introduction}}
\else
\section{Introduction}\label{sec:introduction}
\fi

%
%
%
%


\IEEEPARstart{E}{ach} year, billions of geospatial data points are collected from geolocalized stations (e.g. weather stations), some of which are operated manually and others automatically. These stations are scattered around the world, emit entries at their own rate, and are subject to downtime (e.g. when a natural disaster damages a station's equipment).

Furthermore, some stations (like the ones on board of ships and satellites) are mobile, i.e. their geospatial coordinates change over time. Consequently, geospatial data collected from a large amount and variety of stations is heterogeneous and sparse, making it difficult to organize and further process~\cite{Gandomi2015beyond,Li2016geospatial}.

\subsection{Problem Definition}
Following the aforementioned considerations, the main problem under study is defined as follows: given an arbitrary number $N \in \mathbb{N}_{>0}$ of geospatial observation points of the form $\left(o_\phi \in \left[-90,90\right[;o_\theta \in \left[-180,180\right[;o_\nu \in \mathbb{R}\right)$ scattered on a sphere with radius $\rho \in \mathbb{R}_{>0}$ and an arbitrary number $M \in \mathbb{N}_{>0}$ of geospatial interpolation points of the form $\left(p_\phi \in \left[-90,90\right[;p_\theta \in \left[-180,180\right[\right)$ lying on the same sphere, determine decent interpolation values $p_{\hat{\nu}}$ for each given interpolation point. The problem definition may be visualized as illustrated in \CCReffig{fig:sknni_problem_definition}.

\ShiftFloatSecondCol{%
\CCFigure{fig:sknni_problem_definition}{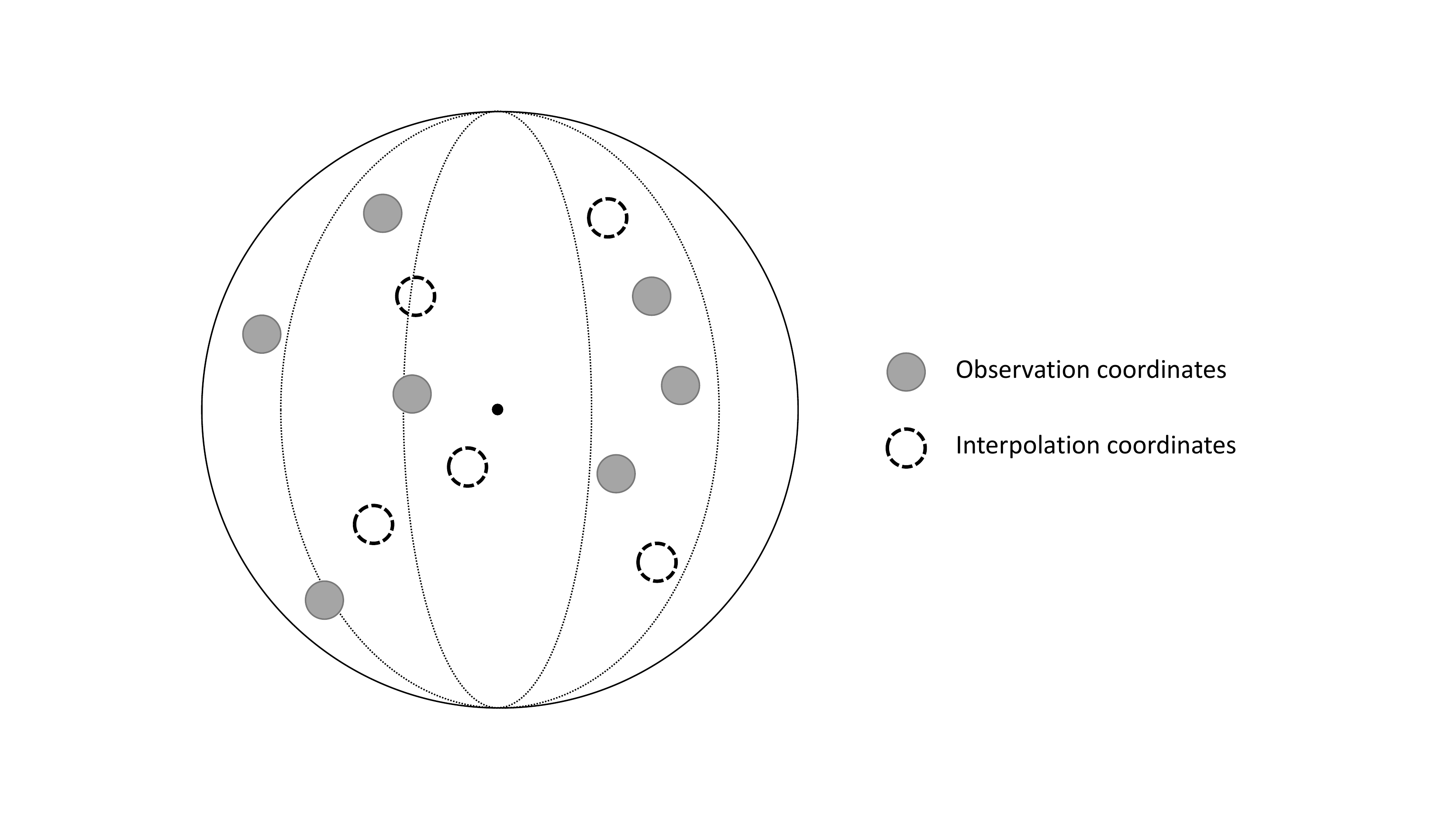}
{Sparse Spherical Interpolation Problem Definition}
{The figure illustrates the problem definition for sparse spherical interpolation. It shows the observation coordinates as filled and the interpolation coordinates as hollow.}
{width=0.85\linewidth,trim=4cm 2cm 4.5cm 2cm,clip}}

\subsection{Motivation}
This work is motivated by the already large and still increasing (by at least \SI{20}{\percent} every year~\cite{Lee2015geospatial}) amount of geospatial data available worldwide, and the need to structure it in a homogeneous way for further processing, e.g. by machine learning and deep learning algorithms and models. Furthermore, geospatial data is useful in a variety of fields like healthcare, security, marketing, environmental modeling, and business intelligence, providing behavioral and evolutional insight~\cite{Jagadish2014big,Erevelles2016big,Chen2014big}. Another motivation for this work is the quality and scope limitations that too many researchers and engineers impose on the geospatial input data used in their projects because there is no simple enough way to work with it and make it useful. This work aims to change that.

\section{Background and Related Work}\label{sec:bg_rw}
Before presenting the algorithm proposed in this work, a review of related algorithms known to perform well for similar tasks is realized and the applicability of each is discussed.

\subsection{Bilinear and Bicubic Interpolation}
When attempting to perform any kind of interpolation, a reoccurring classic is bilinear interpolation and its close relative bicubic interpolation~\cite{Gao2011bilinear}. Put short, bilinear interpolation performs interpolation on a two-dimensional regular grid by first linearly interpolating along one dimension, and then by the other. Its main advantage is being fast to compute, though its main limitation is its interpolation accuracy. Described briefly, bicubic interpolation performs interpolation on a two-dimensional regular grid, using third-order polynomials, by first interpolating along one dimension, and then by the other. Its main advantage is its interpolation accuracy while its main limitation is being more computationally costly. Unfortunately, since both of these algorithms cannot deal with sparse and scattered data points, and especially not on the surface of a sphere, they cannot be retained.

\subsection{Spherical Bivariate Splines}
When attempting to perform spherical interpolation, spherical splines may turn out to be an interesting choice, since they are made to work with spherical data. Their main advantages are applicability and precision, when implemented and tuned correctly. Unfortunately, since this turns out to be challenging in practice, it ends up being one of their principal downsides. Fortunately though, some implementations are available in libraries like SciPy~\cite{Jones2019scipy}, with its smooth sphere bivariate spline~\cite{SciPy2019ssbs} and its LSQ (least-squares) sphere bivariate spline~\cite{SciPy2019lsqsbs}. The main issues with these implementations are weak user friendliness, tricky parameterization, computational expensiveness, and relatively high sensitivity to outliers, problems that turn out to be important enough to also not retain these solutions.

\subsection{Gaussian Process Regression (Kriging)}
Gaussian process regression (also known as Kriging) is a relatively common geostatistical interpolation method which models a phenomenon using a Gaussian process based on prior covariances~\cite{TheTensorflowCommunity2019gaussian}. This approach is interesting since its distribution represents values for a set of interpolation points based on values at a set of observation points, thus being able to work with data not necessarily residing on a regular grid. Though, the main downside of this approach is dealing with ever-changing observation and interpolation coordinates, since a new Gaussian process regression model needs to be created, its kernel chosen and parameters fit, each time a coordinate changes. Such observation and interpolation coordinate changes being very frequent in the context under study make this approach impracticable due to the computational expensiveness required to rebuild and refit the model for each such change.

\section{Proposed Algorithm}\label{sec:main_matter}
Stemming from the motivated need for a simple, flexible and efficient spherical interpolation algorithm, this works proposes \CCSkNNIlong{} (\CCSkNNI{}). The goal of this work is to provide such an algorithm with an interface as simple as possible for ease of use and integration in production pipelines (see \CCReflst{lst:sknni_usage_example} for a usage example), while remaining as flexible as possible to allow more advanced users to customize and adapt the algorithm to their specific (potentially more niche) needs. As such, \CCSkNNI{} is open source and its codebase is available at \CCSknniRepositoryUrl. Thus, anyone interested in customizing the algorithm is welcome to get familiar with the algorithm's inner workings and fork the repository. 

\CCSkNNI{} is a four-part algorithm. The first part consists in performing a change of coordinates to transform the user-space data into a format more practical for the algorithm's inner workings. The second part is about building a spatial index to organize the input observations efficiently. The third part uses the spatial index and transformed interpolation coordinates to find the nearest observation neighbors of each interpolation point. Lastly, the fourth part uses interpolation neighborhood data to estimate the value at each interpolation point. The rest of this section details \CCSkNNI{}'s main execution steps.

\subsection{Change of Coordinate System}
\CCSkNNI{}'s first step is about transforming the observation and interpolation coordinates provided by the user. This is done in two ways: the coordinates are converted from degrees to radians (since all trigonometric operators used in this work operate in radians), and then from polar coordinates to Cartesian coordinates. The transformations are as defined in expressions~\eqref{eq:sknni_tx},~\eqref{eq:sknni_ty} and~\eqref{eq:sknni_tz}.
\begin{align}
\begin{split}
\label{eq:sknni_tx}
\mathcal{T}_x &\colon \mathbb{R}_{\geq-90,<90}
\times \mathbb{R}_{\geq-180,<180}
\times \mathbb{R}_{>0}
\rightarrow \mathbb{R}_{\geq-\rho,\leq\rho}
\end{split}
\\*
\begin{split}
\label{eq:sknni_ty}
\mathcal{T}_y &\colon \mathbb{R}_{\geq-90,<90} 
\times \mathbb{R}_{\geq-180,<180}
\times \mathbb{R}_{>0}
\rightarrow \mathbb{R}_{\geq-\rho,\leq\rho}
\end{split}
\\*
\begin{split}
\label{eq:sknni_tz}
\mathcal{T}_z &\colon \mathbb{R}_{\geq-90,<90}
\times \mathbb{R}_{\geq-180,<180}
\times \mathbb{R}_{>0}
\rightarrow \mathbb{R}_{\geq-\rho,\leq\rho}
\end{split}
\end{align}
The calculations these transformations perform are as described in equations~\eqref{eq:sknni_tx_oi},~\eqref{eq:sknni_ty_oi} and~\eqref{eq:sknni_tz_oi}.
\begin{align}
\small
\begin{split}
\label{eq:sknni_tx_oi}
o_{i,x} &= \mathcal{T}_x\left(o_{i,\phi}, o_{i,\theta}, \rho\right) = 
\rho \cdot 
\cos\left(\frac{\pi \cdot o_{i,\theta}}{180}\right) \cdot 
\sin\left(\frac{\pi \cdot o_{i,\phi}}{180}\right)
\end{split}
\normalsize
\\*
\small
\begin{split}
\label{eq:sknni_ty_oi}
o_{i,y} &= \mathcal{T}_y\left(o_{i,\phi}, o_{i,\theta}, \rho\right) = 
\rho \cdot 
\sin\left(\frac{\pi \cdot o_{i,\theta}}{180}\right) \cdot 
\sin\left(\frac{\pi \cdot o_{i,\phi}}{180}\right)
\end{split}
\normalsize
\\*
\small
\begin{split}
\label{eq:sknni_tz_oi}
o_{i,z} &= \mathcal{T}_z\left(o_{i,\phi}, o_{i,\theta}, \rho\right) = \rho \cdot 
\cos\left(\frac{\pi \cdot o_{i,\phi}}{180}\right)
\end{split}
\normalsize
\end{align}
Here,~$\rho$ corresponds to the user-provided sphere radius, which defaults to the Earth's mean radius of $\SI{6371.01 \pm 0.02}{\kilo\meter}$~\cite{McDonough2003compositional} if not provided. Put short, this step transforms input polar coordinates in degrees into Cartesian coordinates to prepare them for use by further steps.

\subsection{Spatial Index Construction}
\CCSkNNI{}'s second step is about constructing a spatial index to organize the observations provided by the user for efficient querying. The reason why initially provided coordinates are transformed into Cartesian coordinates is to be able to partition the coordinate space. This is done using a $k$\nobreakdash-dimensional tree, as shown in \CCRefeqn{eq:sknni_build_kd_tree}.
\begin{equation}
\label{eq:sknni_build_kd_tree}
\tau = \operatorname{\beta}\left(\left<\left(o_{i,x}, o_{i,y}, o_{i,z}\right)\right>_{\forall i \in \left\{1, 2, \ldots, N\right\}}\right)
\end{equation}
Here, $\beta$ reprensents the $k$\nobreakdash-dimensional tree building operator, angle brackets ($\left<\right>$) represent a list of elements, and $\tau$ represents the built $k$\nobreakdash-dimensional tree (in this case, a 3\nobreakdash-dimensional tree). Since an efficient implementation of $k$\nobreakdash-dimensional trees is already present in SciPy (see~\cite{Scipy2019ckdtree}), this work simply uses it as is. Succinctly, this step organizes observations by building a spatial index (a $k$\nobreakdash-dimensional tree) to allow for efficient querying of nearest neighbors in further steps.

\subsection{Neighbor Finding}
\CCSkNNI{}'s third step is about finding the $k$\nobreakdash-nearest observation neighbors of each interpolation point, since information about these nearest neighbors will be used to estimate the value at each interpolation point. To find the $k$\nobreakdash-nearest observation neighbors of each interpolation point, a query is issued to the $k$\nobreakdash-dimensional tree built earlier. One of the nice features of SciPy's $k$\nobreakdash-dimensional tree implementation is that they accept batch queries, so the queries for the $k$\nobreakdash-nearest observation neighbors of all interpolation points can be computed at once (facilitating hardware and software acceleration). This batch querying is as shown in equations~\eqref{eq:sknni_query_kd_tree_query} and~\eqref{eq:sknni_query_kd_tree_query_result}.
\begin{IEEEeqnarray}{cCl}
	\label{eq:sknni_query_kd_tree_query}
	\mathfrak{N} &= \tau\left(\left[
	\begin{IEEEeqnarraybox*}[][c]{,c/c/c,}
		p_{1,x} & p_{1,y} & p_{1,z}\\
		p_{2,x} & p_{2,y} & p_{2,z}\\
		\vdots & \vdots & \vdots\\
		p_{M,x} & p_{M,y} & p_{M,z}
	\end{IEEEeqnarraybox*}
	\right]\right) \\
	\label{eq:sknni_query_kd_tree_query_result}
	&= \left[
	\begin{IEEEeqnarraybox*}[][c]{,c/c/c/c,}
		\mathfrak{N}_{1,1} & \mathfrak{N}_{1,2} & \ldots & \mathfrak{N}_{1,k}\\
		\mathfrak{N}_{2,1} & \mathfrak{N}_{2,2} & \ldots & \mathfrak{N}_{2,k}\\
		\vdots & \vdots & \ddots & \vdots\\
		\mathfrak{N}_{M,1} & \mathfrak{N}_{M,2} & \ldots & \mathfrak{N}_{M,k}
	\end{IEEEeqnarraybox*}
	\right]
\end{IEEEeqnarray}

The query result is a matrix of which each row represents an interpolation point's neighborhood (in proximity order). The columns represent the proximity order, i.e. column~\num{1} represents the closest neighboring observation, column~\num{2} represents the second closest neighboring observation, and so on. Each value of the matrix is an index corresponding to the obseration point's index in the observation list that was provided when building the $k$\nobreakdash-dimensional tree. With all these indices, it is now possible to determine each neighbor's information simply by indexing the observation list. Since the matrix can be represented programmatically as a NumPy index array~\cite{SciPy2019indexing}, this multi-indexing can be done at once, as represented in \CCRefeqn{eq:sknni_neighb_obs_indexing}.
\begin{IEEEeqnarray}{cCl}
	\label{eq:sknni_neighb_obs_indexing}
	O_\mathfrak{N} = \left[
	\begin{IEEEeqnarraybox*}[][c]{,c/c/c/c,}
		o_{\mathfrak{N}_{1,1}} & o_{\mathfrak{N}_{1,2}} & \ldots & o_{\mathfrak{N}_{1,k}}\\
		o_{\mathfrak{N}_{2,1}} & o_{\mathfrak{N}_{2,2}} & \ldots & o_{\mathfrak{N}_{2,k}}\\
		\vdots & \vdots & \ddots & \vdots\\
		o_{\mathfrak{N}_{M,1}} & o_{\mathfrak{N}_{M,2}} & \ldots & o_{\mathfrak{N}_{M,k}}
	\end{IEEEeqnarraybox*}
	\right]
\end{IEEEeqnarray}
Put simply, this step finds information about the $k$\nobreakdash-nearest observation neighbors of each interpolation point.

\subsection{Interpolation}
\CCSkNNI{}'s fourth and last step is about estimating the value at each interpolation point based on its observation neighborhood. To do so, an interpolation function of the form defined in \CCRefexp{eq:sknni_interp_fn_def} is used. The specific calculations executed during this step are determined by the chosen interpolation function, which is called as shown in \CCRefeqn{eq:sknni_interp_fn_generic} to obtain the estimated (interpolated) values at each interpolation point, in order.
\begin{align}
\begin{split}
\label{eq:sknni_interp_fn_def}
\mathfrak{I} : \mathbb{R}_{\geq-\frac{\pi}{2},<\frac{\pi}{2}}^M \times \mathbb{R}_{\geq-\pi,<\pi}^M \times \mathbb{R}_{\geq-\frac{\pi}{2},<\frac{\pi}{2}}^{M \times k} \times \mathbb{R}_{\geq-\pi,<\pi}^{M \times k} \\
\phantom\space \times \mathbb{R}^{M \times k} \times \mathbb{R}_{>0} \times \mathbb{N}_{>0,\leq N} \rightarrow \mathbb{R}^M
\end{split}
\\*
\begin{split}
\label{eq:sknni_interp_fn_generic}
P_{\hat\nu} = \mathfrak{I}\left(O_{\mathfrak{N}, \phi}, O_{\mathfrak{N}, \theta}, O_{\mathfrak{N}, \nu}, P_\phi, P_\theta, \rho, k\right)
\end{split}
\end{align}

Note that \CCSkNNI{} provides a default interpolation function (see \CCRefsec{sec:nddnisd}) that performs well in many synthetic and real world cases (see \CCRefsec{sec:results_analysis}). In summary, this last step passes interpolation points with their observation neighborhood information to an interpolation function to estimate the value at each interpolation point. The estimated values are then associated with their respective user-space-format interpolation points, and the overall result is returned to user space.

\section{NDDNISD Interpolation Function}\label{sec:nddnisd}
\CCNDDNISDlong{} (\CCNDDNISD{}) is \CCSkNNI{}'s default interpolation function. It is comprised of four main steps. The first one is about calculating the distance between each interpolation point and each observation of its neighborhood. The second step computes proximal weights based on the aforementioned distances. The third step performs neighborhood distribution debiasing (see \CCRefsec{sec:nddnisd_ndd} for more details) based on prior proximal weights and neighborhood spatial distribution. The fourth and last step is about integrating the insight gained throughout the algorithm's execution and using it to produce the estimated values at each interpolation point.

\subsection{Proximity Assessment}\label{sec:nddnisd_orth_dist}
\CCNDDNISD{}'s first step is about proximity assessment, i.e. the goal is to determine the orthodromic distance (the minimal distance, on the surface of a sphere, between two points~\cite{WolframReaearchInc2019great}, as shown in \CCReffig{fig:sknni_orthodromic_distance}) between each interpolation point and each of its observation neighbors. These orthodromic distances turn out to be important since they are a good heuristic to estimate the relevance of each neighborhood's observations.

\CCFigure{fig:sknni_orthodromic_distance}{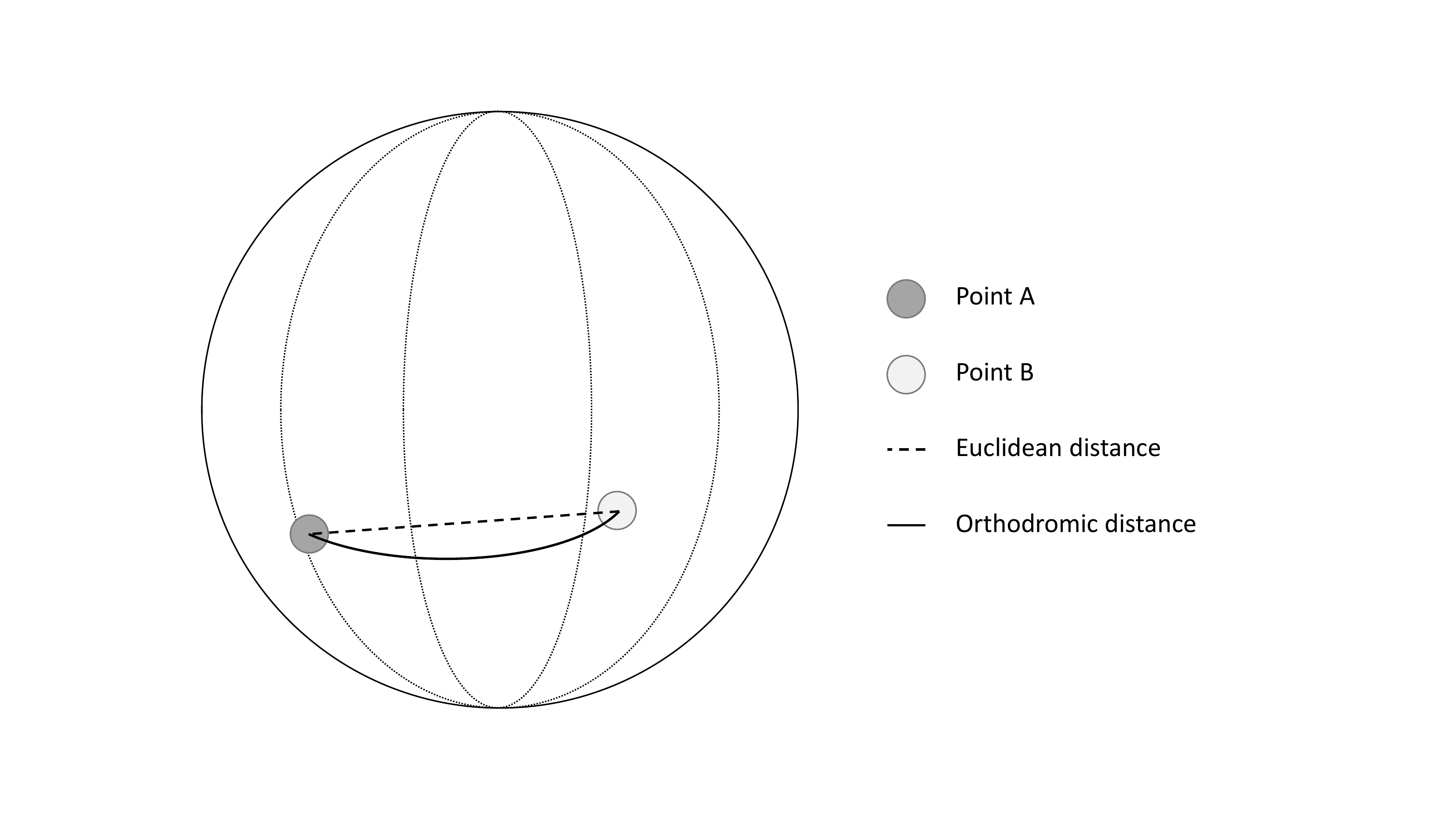}
{Orthodromic Distance}
{The figure shows the difference between orthodromic distance and Euclidean distance between two points, $A$ and $B$, on the surface of a sphere.}
{width=0.85\linewidth,trim=3cm 2cm 3cm 2cm,clip}

Since orthodromic distances are on the surface of a sphere, calculating them is more complex than calculating Euclidean distances. Orthodromic distance calculation is as defined in expressions~\eqref{eq:sknni_great_circle_dist_a},~\eqref{eq:sknni_great_circle_dist_a_num_stability}~and~\eqref{eq:sknni_great_circle_dist_d}.
\begin{align}
\footnotesize
\begin{split}
\label{eq:sknni_great_circle_dist_a}
&a \leftarrow \operatorname{\sin^2}\left(\frac{B_\phi - A_\phi}{2}\right) 
+ \cos\left(A_\phi\right)\cos\left(B_\phi\right)
\operatorname{\sin^2}\left(\frac{B_\theta - A_\theta}{2}\right)
\end{split}
\normalsize
\\*
\footnotesize
\begin{split}
\label{eq:sknni_great_circle_dist_a_num_stability}
&a \leftarrow \min\left(1,\max\left(0,a\right)\right)
\end{split}
\normalsize
\\*
\footnotesize
\begin{split}
\label{eq:sknni_great_circle_dist_d}
&\mathfrak{D}\left(A_\phi, A_\theta, B_\phi, B_\theta, \rho\right) = 
2 \rho \cdot \operatorname{arctan2}\left(\sqrt{a}, \sqrt{1 - a}\right)
\end{split}
\normalsize
\end{align}
Here, \CCRefexp{eq:sknni_great_circle_dist_a} makes use of the haversine formula to calculate $a$, which corresponds to the square of half the chord length between the two points (as if they were on the unit sphere)~\cite{Veness2019calculate}. Since precautions regarding numerical stability must be taken when implementing orthodromic distance calculation, \CCRefexp{eq:sknni_great_circle_dist_a_num_stability} ensures numerical stability by clipping $a$ such that $a \in \left[0;1\right]$. Then, \CCRefexp{eq:sknni_great_circle_dist_d} determines the arc length in radians between the two points and uses it in conjunction with the sphere's radius~$\rho$ to end up with the orthodromic distance between the two points. 

Since these calculations can be applied element-wise on vectors, matrices and collections of higher dimensionality, the implementation leverages NumPy array broadcasting and element-wise operations to compute all orthodromic distances at once, allowing for further hardware and software optimization.

\subsection{Normalized Inverse Squared Distance}\label{sec:nddnisd_nisd}
\CCNDDNISD{}'s second step is about proximal weighting, i.e. the aim is to determine each observation neighbor's relevance based on its orthodromic distance to the interpolation point of interest, as shown in \CCReffig{fig:sknni_neighborhood}. 
\CCFigure{fig:sknni_neighborhood}{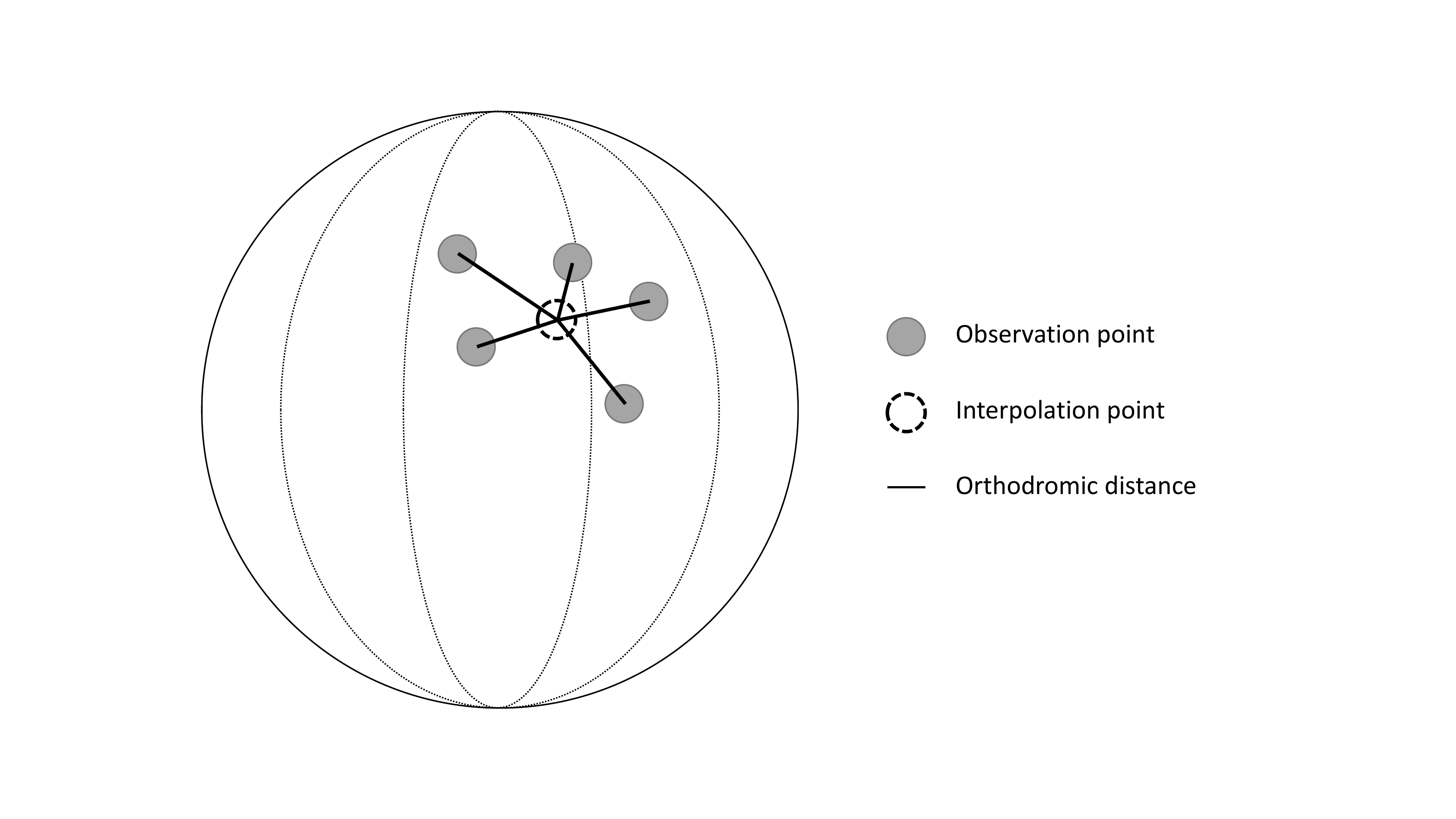}
{Interpolation Neighborhood}
{The figure shows an interpolation neighborhood with the orthodromic distances between its interpolation coordinates and the ones of each observation neighbor. These distances are the ones used to determine the importance of each neighbor.}
{width=0.85\linewidth,trim=3cm 2cm 3cm 2cm,clip}
As such, \CCNDDNISD{} uses proximal weights calculated by \CCNISDlong{} (\CCNISD{}) weighting as shown in \CCRefeqn{eq:sknni_nisd}.
\begin{equation}
\label{eq:sknni_nisd}
\operatorname{NIS}\left(d\right) = 
\frac{\frac{1}{d^2+\varepsilon}}
{\sum_{i=1}^k{\frac{1}{d_i^2+\varepsilon}}}
\end{equation}
As its name implies, this function takes in distances, squares them, adds a very small strictly positive constant $\varepsilon \rightarrow 0^+$ to them to ensure numerical stability, inverts them, and normalizes them by dividing them by their sum. The resulting values are the proximal weights which are passed on to the next step.

\subsection{Neighborhood Distribution Debiasing}\label{sec:nddnisd_ndd}
\CCNDDNISD{}'s third step consists in performing \CCNDDlong{} (\CCNDD{}). This step acts as a reweighting of the proximal weights based on the bias of the spatial distribution of observation neighbors. The main idea behind \CCNDD{} is illustrated in \CCReffig{fig:sknni_ndd} and its execution details are as follows.

\CCFigure{fig:sknni_ndd}{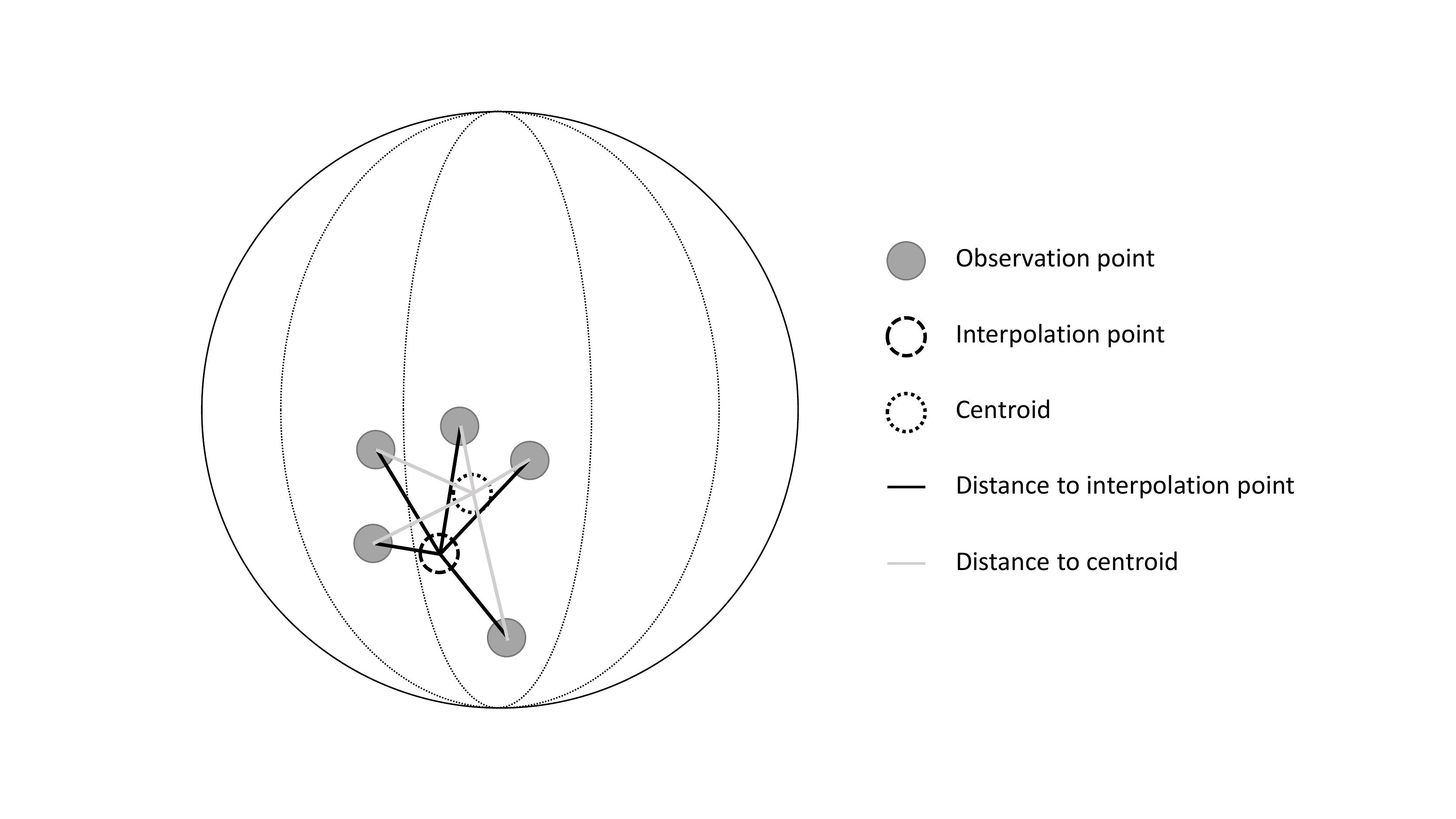}
{Neighborhood Distribution Debiasing}
{The figure shows a biased interpolation neighborhood. The distances from observations to the interpolation point are also illustrated as well as distances from observations to the neighborhood's centroid. Intuitively, the more an observation neighbor is close to the neighborhood's centroid, the more it is biased. NDD uses this information to renormalize the proximal weights.}
{width=0.85\linewidth,trim=3cm 2cm 3cm 2cm,clip}

Starting with orthodromic distances~$\delta$ obtained as shown in \CCRefeqn{eq:sknni_ndd_orth_dist} (see \CCRefsec{sec:nddnisd_orth_dist}) which are used to compute proximal weights~$w_\delta$ as shown in \CCRefeqn{eq:sknni_ndd_prox_w} (see \CCRefsec{sec:nddnisd_nisd}), \CCNDD{} first determines the observation neighborhood's centroid.
\begin{align}
\begin{split}
\label{eq:sknni_ndd_orth_dist}
\delta = \mathfrak{D}\left(P_\phi, P_\theta, O_{\mathfrak{N}, \phi}, O_{\mathfrak{N}, \theta}, \rho\right)
\end{split}
\\
\begin{split}
\label{eq:sknni_ndd_prox_w}
w_\delta = \operatorname{NIS}\left(\delta\right)
\end{split}
\end{align}

Calculating the observation neighborhood's centroid is relatively straightforward: its latitudinal component is calculated as shown in \CCRefeqn{eq:sknni_ndd_centroid_lat} and its longitudinal component is calculated as shown in \CCRefeqn{eq:sknni_ndd_centroid_lng}.
\begin{align}
\begin{split}
\label{eq:sknni_ndd_centroid_lat}
\widebar{\phi} = \frac{1}{k}\sum_{i=1}^{k}{O_{\mathfrak{N}, \phi, i}}
\end{split}
\\
\begin{split}
\label{eq:sknni_ndd_centroid_lng}
\widebar{\theta} = \frac{1}{k}\sum_{i=1}^{k}{O_{\mathfrak{N}, \theta, i}}
\end{split}
\end{align}
The centroid is thus defined as the $(\widebar{\phi},\widebar{\theta})$ point. Such centroid is calculated for each interpolation neighborhood. Then, the orthodromic distance between each observation neighbor and its neighborhood's centroid (see \CCReffig{fig:sknni_ndd}) is calculated as shown in \CCRefeqn{eq:sknni_ndd_centroid_dist}.
\begin{align}
\begin{split}
\label{eq:sknni_ndd_centroid_dist}
\eta = \mathfrak{D}\left(\widebar{\phi}, \widebar{\theta}, O_{\mathfrak{N}, \phi}, O_{\mathfrak{N}, \theta}, \rho\right)
\end{split}
\end{align}

Lastly, the centroidal distances~$\eta$ are used to renormalize the prior proximal weights~$w_\delta$ as shown in \CCRefeqn{eq:sknni_ndd_ndd_w}.
\begin{align}
\begin{split}
\label{eq:sknni_ndd_ndd_w}
w_\eta = \frac{w_\delta \eta}{\sum_{i=1}^k{w_{\delta, i} \eta_i}}
\end{split}
\end{align}

This renormalization effectively reduces the weight of observation neighbors based on how biased (close to their neighborhood's centroid) they are. Thus, this step results in neighborhood-distribution-debiased proximal weights~$w_\eta$ which are passed further on for integration.

\subsection{Integration}\label{sec:nddnisd_integration}
\CCNDDNISD{}'s fourth and last step simply integrates the insight gathered throughout by computing, for each interpolation neighborhood, a weighted sum of its neiboring observation values~$O_{\mathfrak{N}, \nu}$ and their associated neighborhood-distribution-debiased proximal weights~$w_\eta$, as shown in \CCRefeqn{eq:sknni_ndd_vals}.
\begin{align}
\begin{split}
\label{eq:sknni_ndd_vals}
P_{\hat{\nu}} = \sum_{i=1}^{k}{w_{\eta} O_{\mathfrak{N}, \nu}}
\end{split}
\end{align}
The output of the integration step is a list containing the estimated value at each requested interpolation point.

\section{Algorithm Evaluation}\label{sec:evaluation}
\CCSkNNI{}'s uniqueness makes it tricky to evaluate, since there is no analogous algorithm to compare it to. Thus, its default interpolation function~$\mathfrak{I}_{\text{\CCNDDNISD{}}}$ is evaluated by comparing it to meaningful baselines. 

The first baseline interpolation function is the $\mathfrak{I}_{\text{Nearest}}$ function (defined in \CCRefeqn{eq:sknni_interp_fn_nearest}) which only considers the nearest neighbor of the neighborhood and disregards the rest.
\begin{align}
\footnotesize
\begin{split}
\label{eq:sknni_interp_fn_nearest}
\mathfrak{I}_{\text{Nearest}}\left(O_{\mathfrak{N}, \phi}, O_{\mathfrak{N}, \theta}, O_{\mathfrak{N}, \nu}, P_\phi, P_\theta, \rho, k\right)
&= \frac{1}{k}\sum_{i=1}^{k}{\mathds{1}_{i=1} O_{\mathfrak{N}_i, \nu}}
\end{split}
\normalsize
\end{align}
Here, $\mathds{1}$ is an indicator function which takes the value \num{1} if its condition is true and \num{0} otherwise. This interpolation function is indeed quite simple, but is considered in the algorithm's evaluation process due to its widespread and surprisingly frequent use in many industrial applications.

The second comparison baseline is the $\mathfrak{I}_{\text{Mean}}$ interpolation function (defined in \CCRefeqn{eq:sknni_interp_fn_mean}) which calculates the neighborhood's mean value.
\begin{align}
\footnotesize
\begin{split}
\label{eq:sknni_interp_fn_mean}
\mathfrak{I}_{\text{Mean}}\left(O_{\mathfrak{N}, \phi}, O_{\mathfrak{N}, \theta}, O_{\mathfrak{N}, \nu}, P_\phi, P_\theta, \rho, k\right)
&= \frac{1}{k}\sum_{i=1}^{k}{O_{\mathfrak{N}_i, \nu}}
\end{split}
\normalsize
\end{align}
This function is considered since calculating a group's mean value is a common reduction function.

The third comparison baseline is the $\mathfrak{I}_{\text{Median}}$ interpolation function (defined in \CCRefeqn{eq:sknni_interp_fn_median}) which calculates the neighborhood's median value.
\begin{align}
\footnotesize
\begin{split}
\label{eq:sknni_interp_fn_median}
\mathfrak{I}_{\text{Median}}\left(O_{\mathfrak{N}, \phi}, O_{\mathfrak{N}, \theta}, O_{\mathfrak{N}, \nu}, P_\phi, P_\theta, \rho, k\right)
&= \operatorname{median}\left(O_{\mathfrak{N}_i, \nu}\right)
\end{split}
\normalsize
\end{align}
This function is also considered since the median is also a common reduction function which is often less sensitive to outliers than the mean.

With the interpolation functions to compare now defined, the evaluation process is as described in \CCReffig{fig:sknni_evaluation}. This evaluation process is repeated for each quantity nature~$\Upsilon$, for each run~$R \in \{1,2,\ldots,100\}$ of the same experiment with a new observation set, for each interpolation function~$\mathfrak{I}$, for each number of nearest neighbors to consider $k \in \{1,2,\ldots,25\}$. The quantity natures cover both synthetic data ($\Upsilon_{\text{Synthetic}}$, for comparison with theoretical predictions) and real world data ($\{\Upsilon_{\text{Temperature}},\Upsilon_{\text{DewPoint}},\Upsilon_{\text{Pressure}},\Upsilon_{\text{WindSpeed}}\}$, for applicability and practicality assessment). The observation sets used contain about \num{4000} observations each, \num{1000} of which are randomly sampled without replacement to build the interpolator while the rest becomes the holdout set. Each $(\mathfrak{I},k)$ configuration then estimates the value at each hidden point and the results are collected for further analysis.

\CCWideFigure{fig:sknni_evaluation}{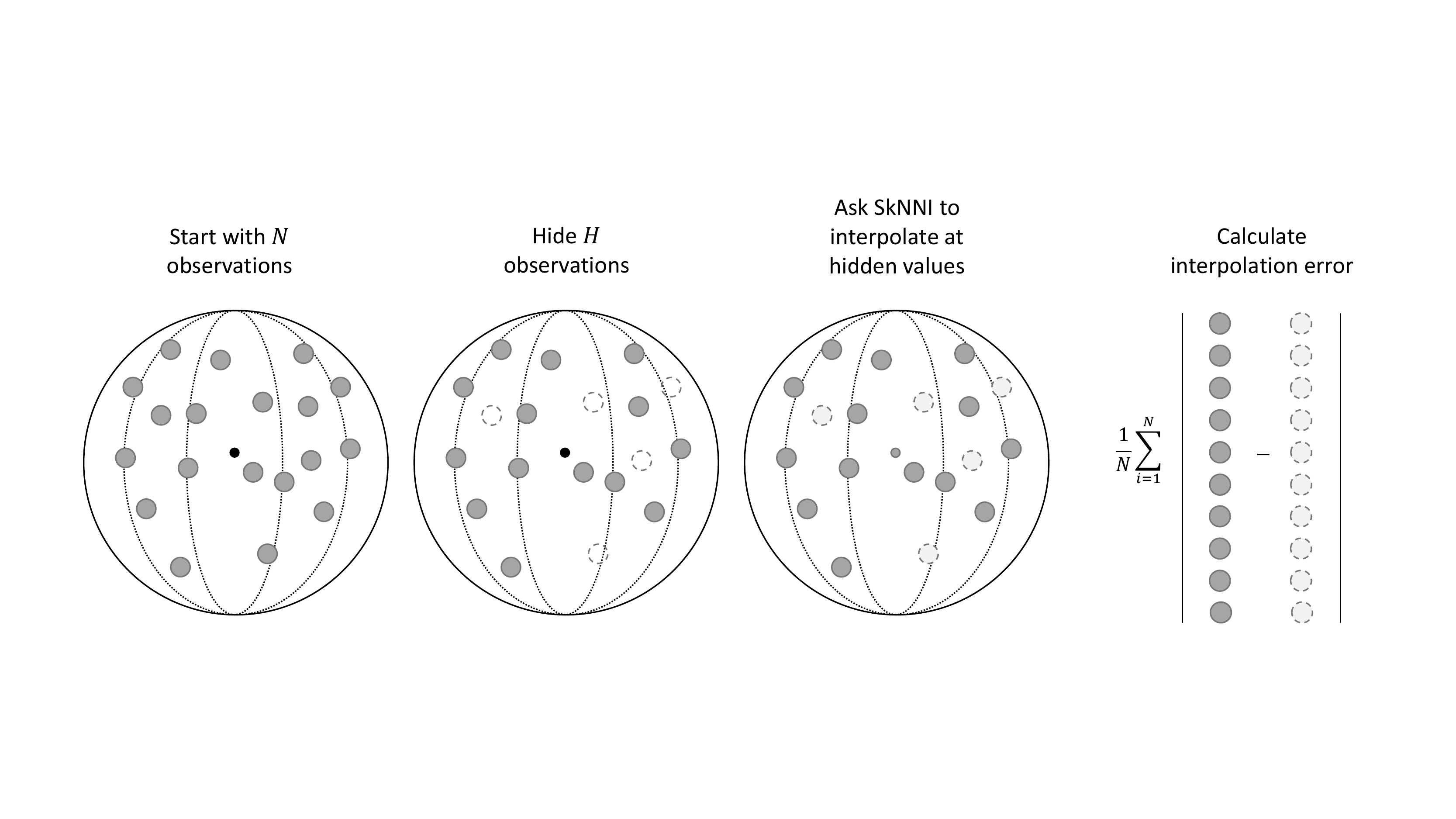}
{Evaluation Process for SkNNI}
{The figure shows the evaluation process for SkNNI, which is performed in four main steps. The first step consists in acquiring a set of observation values. The second step then hides a portion of the observations in a holdout set used for evaluation. The third step consists in having SkNNI perform interpolation at the coordinates where values were hidden. The fourth and last step then simply measures the algorithm's accuracy using a given metric.}
{width=0.85\linewidth,trim=1cm 4cm 1cm 4cm,clip}

\subsection{AMERPE Metric}
To compare the interpolation functions~$\mathfrak{I}$ fairly on various quantity natures~$\Upsilon$, this work defines a custom metric: the \CCAMERPElong{} (\CCAMERPE{}), which is as defined in \CCRefeqn{eq:sknni_eval_amerpe}.
\begin{align}
\begin{split}
\label{eq:sknni_eval_amerpe}
\operatorname{AMERPE}\left(\widetilde{\mathcal{V}},\widehat{\mathcal{V}};\mathcal{V}_{\text{min}},\mathcal{V}_{\text{max}}\right) = \frac{100}{\mathcal{V}_{\text{max}} - \mathcal{V}_{\text{min}}} \left|\widetilde{\mathcal{V}} - \widehat{\mathcal{V}}\right|
\end{split}
\end{align}
The idea behind \CCAMERPE{} is to first calculate the absolute error between the observed (true) value~$\widetilde{\mathcal{V}}$ and the predicted value~$\widehat{\mathcal{V}}$. Then, the calculated absolute prediction error is divided by the variation range of the quantity nature which is considered to have the domain $[\mathcal{V}_{\text{min}};\mathcal{V}_{\text{max}}]$. This is done to determine the ratio of the absolute prediction error with respect to the maximal prediction error that could ever be made. This ratio is then simply converted into a percentage by multiplying it by \num{100}. The resulting quantity is the \CCAMERPE{}, a quantity expressed without units that represents how inaccurate a predictor is with respect to how inaccurate it could ever be, on a scale from \num{0} (minimal error) to \num{100} (maximal error).

To illustrate the idea and show why \CCAMERPE{} is useful, consider the following example. Let the observed (true) value (that is kept hidden during experimentation) be \num{2} and the predicted value be \num{3}. This yields an absolute interpolation error of $\left|2 - 3\right| = 1$. If the quantity nature is known to vary from \num{-10} to \num{40}, its variation range is $40 - -10 = 50$. Thus the \CCAMERPE{} is $100 \times 1 \div 50 = 2$. This can be interpreted as the interpolator making an error that is only \SI{2}{\percent} of the maximal error it could have ever made, which is quite good. Now, if the same absolute interpolation error of $\left|2 - 3\right| = 1$ occurs, but for a quantity nature that is instead known to vary from \num{-1} to \num{4}, the variation range becomes $4 - -1 = 5$. In that case, the \CCAMERPE{} becomes $100 \times 1 \div 5 = 20$, which can be interpreted as the interpolator making an error that is \SI{20}{\percent} of the maximal error it could have ever made, which is \num{10} times as bad as making the same absolute error on the former quantity nature.

In summary, \CCAMERPE{} is a metric that relativizes absolute error with respect to the maximal error that could ever be made, which allows for fairer comparison of interpolators on various quantity natures.

\subsection{Evaluation Data}\label{sec:evaluation_eval_data}
In this work, evaluation data is separated in two main categories: synthetic data, which is generated by a known synthetic geospatial function, and real world data which comes from the sensors of real world stations.

The synthetic geospatial function used in this work is as defined in expressions~\eqref{eq:sknni_synthetic_lat} through~\eqref{eq:sknni_synthetic_gen_fn}.
\begin{align}
\begin{split}
\label{eq:sknni_synthetic_lat}
\Phi \sim \mathcal{C}\left(0, 30, -90, 90\right)
\end{split}
\\*
\begin{split}
\label{eq:sknni_synthetic_lng1}
\Theta_1 \sim \mathcal{C}\left(0, 60, -180, 180\right)
\end{split}
\\*
\begin{split}
\label{eq:sknni_synthetic_lng2}
\Theta_2 \sim \mathcal{S}\left(
\left<-125, -75, 0, 75, 100, 135\right>,\right.\\
\left.
\left<0.15, 0.15, 0.15, 0.2, 0.2, 0.15\right>
\right)
\end{split}
\\*
\begin{split}
\label{eq:sknni_synthetic_lng}
\Theta = (\Theta_1 + \Theta_2 + 180) \operatorname{mod}^+ 360 - 180
\end{split}
\\*
\begin{split}
\label{eq:sknni_synthetic_noise}
\mathfrak{Z} \sim \mathcal{U}\left(0, 8\right)
\end{split}
\\*
\begin{split}
\label{eq:sknni_synthetic_gen_fn}
\mathcal{V} &= \mathcal{G}\left(\Phi, \Theta, \mathfrak{T}, \mathfrak{Z}\right) = 42 \sin\left(\frac{\pi(\Phi + 90)}{180}\right) \\
\phantom\space &+ 7 \cos\left(\frac{3}{2}\frac{\pi(\Theta + 180)}{180} + \frac{\pi}{12}\mathfrak{T}\right) + \mathfrak{Z} - 25
\end{split}
\end{align}
Here, $\mathcal{C}\left(\mu,\sigma,a,b\right)$ represents a truncated normal distribution, $\mathcal{U}\left(a,b\right)$ represents a uniform distribution, $\mathcal{S}\left(A,P\right)$ represents random sampling with replacement from elements of $A$ with probabilities $P$ where the probability of selecting $a_i \in A$ is $p_i \in P \,| \sum_{p_i \in P}{p_i} = 1$, and $\operatorname{mod}^+$ represents the modulo operator which returns the first positive remainder.

As such, each virtual station's latitude is sampled from the distribution shown in \CCRefexp{eq:sknni_synthetic_lat} and each virtual station's longitude is calculated using \CCRefeqn{eq:sknni_synthetic_lng} based on samples from the distributions shown in expressions~\eqref{eq:sknni_synthetic_lng1} and~\eqref{eq:sknni_synthetic_lng2}. The synthetic observations are then calculated using \CCRefeqn{eq:sknni_synthetic_gen_fn} based on latitude, longitude, time and a noise term which samples its values from the distribution shown in \CCRefexp{eq:sknni_synthetic_noise}. 

This formulation allows the use of different time~$\mathfrak{T}$ values to generate each experiment's observation set, making~$\mathcal{G}$ behave like a real noisy geospatial function evolving through time. Furthermore, an important effect of the uniform noise~$\mathfrak{Z}$ is that it enforces a limit on the expected minimal absolute interpolation error. This theoretical limit is as described in \CCRefeqn{eq:sknni_exp_min_abs_err_lim}, assuming~$\mathcal{V}$ follows a uniform distribution~$\mathcal{U}\left(a,b\right)$ where $a$ is the uniform distribution's lower bound and $b$ its upper bound (see~\cite{Trempe2019worldwide} for the lemma and associated proof).
\begin{align}
\begin{split}
\label{eq:sknni_exp_min_abs_err_lim}
\mathbb{E}\left[\left|\widetilde{\mathcal{V}} - \mathbb{E}\left[\mathcal{V}\right]\right|\right] = \frac{b\,-\,a}{4}
\end{split}
\end{align}

Thus, using the $a$ and $b$ values of the noise~$\mathfrak{Z}$ distribution and substituting them in \CCRefeqn{eq:sknni_exp_min_abs_err_lim} yields an unavoidable expected interpolation error of $(8 - 0) \div 4 = 2$. Analyzing the synthetic function~$\mathcal{G}$, its extrema are $\min\left(\mathcal{G}\right) = -32$ and $\max\left(\mathcal{G}\right) = 32$. When converted into \CCAMERPE{} using the synthetic function's extrema, the expected minimal \CCAMERPE{} is $100 \times 2 \div (32 - -32) = 3.125$. This means that, over a large number of experiments, the theoretically minimal \CCAMERPE{} achievable when interpolating for the synthetic function~$\mathcal{G}$ is \num{3.125} because of the uniform noise~$\mathfrak{Z}$ term present in the calculation of its values.

Now, the real world data used in this work's experiments is much simpler to describe since it consists of real world observations originating from station sensors and recording devices. The real world data used in this work is hourly-aggregated worldwide weather observation data for temperature, dew point, pressure and wind speed, data which was graciously provided by \CCPelmorexCorp{}

\section{Empirical Results and Analysis}\label{sec:results_analysis}
After running all the experimental configurations detailed in \CCRefsec{sec:evaluation}, results are first aggregated by quantity nature~$\Upsilon$ (for each figure), and then by interpolation function~$\mathfrak{I}$ and number of nearest neighbors~$k$ (for every bar in the figures). The methodology ensures high statistical significance by calculating statistics on \num{300000} truth-interpolation pairs per bar of every figure, for a total of \num{30000000} truth-interpolation pairs per figure. This section now presents and discusses the results for interpolation on synthetic and real world data.

Starting with interpolation on synthetic geospatial data, the results are as shown in \CCReffig{fig:sknni_res_synthetic}.
\CCFigure{fig:sknni_res_synthetic}{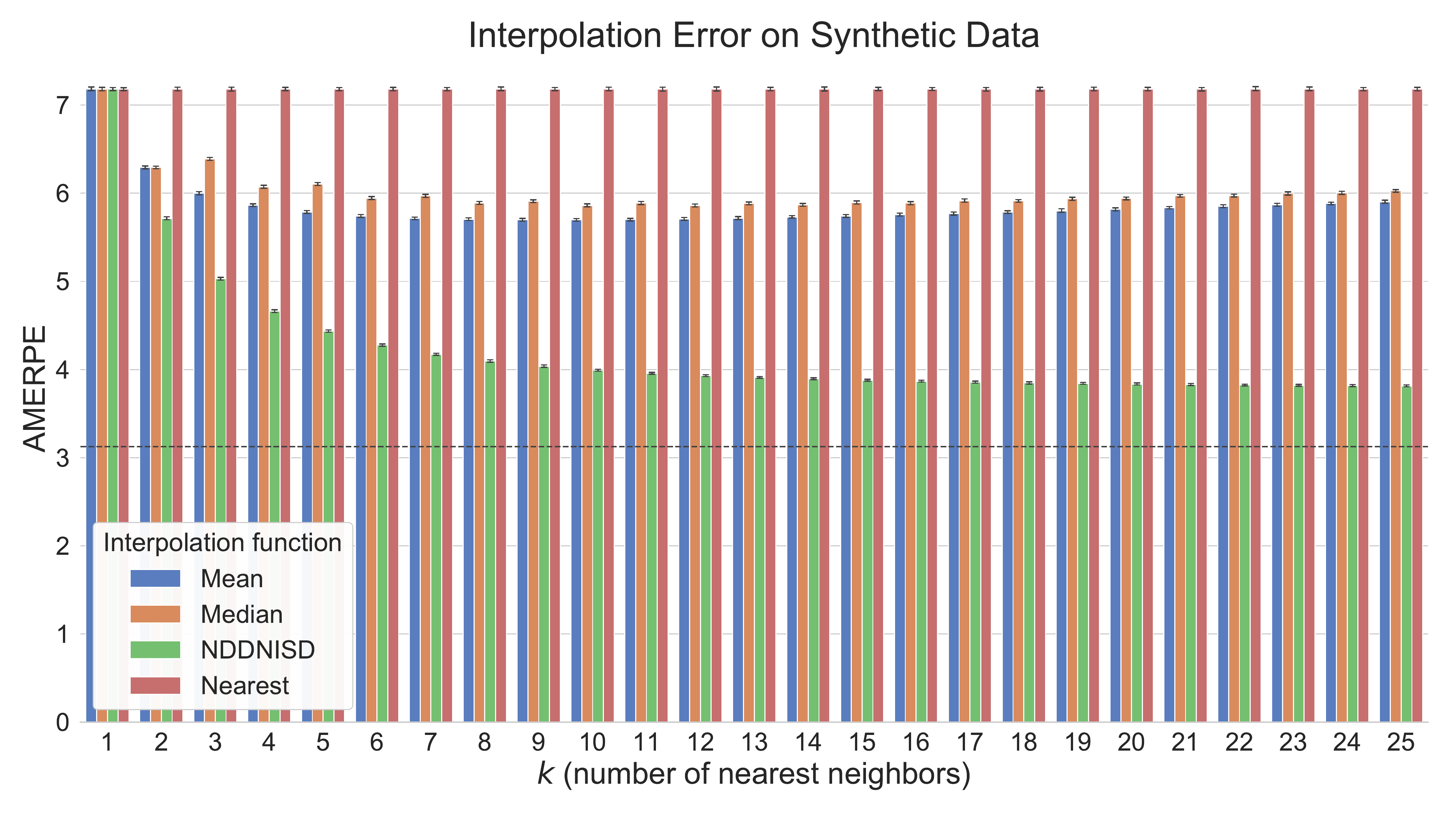}
{Interpolation Error on Synthetic Data}
{The figure shows the interpolation error of various interpolation functions on noisy synthetic data. \num{100} experiments were run, each for a different variant of the noisy geospatial function~$\mathcal{G}$, and AMERPE was calculated for various interpolation functions for \num{25} values of $k$ (number of nearest neighbors) using \num{1000} observations and \num{3000} validation observations that were held out for evaluation. The error bars represent the \SI{95}{\percent} confidence interval for the bootstrap mean (over \num{100} bootstrap samples) of the evaluated quantities. The dashed horizontal line represents the best (minimal) error expectation considering the noise present in the geospatial function.}%
{width=0.9\linewidth}
At first glance, $\mathfrak{I}_{\text{Nearest}}$ appears to perform the worst. It also does not improve when more neighbors are considered (as expected) since it always selects only the value of the closest neighbor. Nonetheless, considering $\mathfrak{I}_{\text{Nearest}}$ is relevant as will be discussed further on. $\mathfrak{I}_{\text{Mean}}$ and $\mathfrak{I}_{\text{Median}}$ perform similarly as their error diminishes when increasing the number of considered nearest neighbors~$k$, up to a point after which error starts to go up, even though very slightly. This is likely explained by the further neighbors acting more like noise than useful information. 

Besides, $\mathfrak{I}_{\text{\CCNDDNISD{}}}$ significantly outperforms the other interpolation functions, achieving an interpolation error about \num{4} times closer to the theoretical limit of \num{3.125} (as discussed in \CCRefsec{sec:evaluation_eval_data}) than the second-best interpolation function, in this case $\mathfrak{I}_{\text{Mean}}$. Furthermore, $\mathfrak{I}_{\text{\CCNDDNISD{}}}$'s error does not seem to start increasing after any number of nearest neighbors to consider~$k$. This result might be interpreted as $\mathfrak{I}_{\text{\CCNDDNISD{}}}$ being capable of extracting useful information from each additional neighbor, even further ones, while also remaining robust to noise in observation values.

Moving on to interpolation on real world data, results are presented in \CCReffig{fig:sknni_res_rw}. A first element to note is that $\mathfrak{I}_{\text{Nearest}}$ can now outperform both $\mathfrak{I}_{\text{Mean}}$ and $\mathfrak{I}_{\text{Median}}$ as shown in subfigures \ref{fig:interp_rw_temperature}, \ref{fig:interp_rw_dew_point} and \ref{fig:interp_rw_pressure}. A likely explanation is that the nature of such real world data is very localized in space, meaning only observations very close to the interpolation point are relevant while others cumulate to a large amount of noise. This large amount of noisy further observations would therefore affect $\mathfrak{I}_{\text{Mean}}$ and $\mathfrak{I}_{\text{Median}}$, but not $\mathfrak{I}_{\text{Nearest}}$ since it only considers the closest (and potentially most meaningful) neighboring observation.

Altough, an interesting case is for interpolation on wind speed data (see \CCReffig{fig:interp_rw_wind_speed}), where interpolators perform analogously to how they performed on synthetic data (see \CCReffig{fig:sknni_res_synthetic}). In these cases, it is likely the data's nature makes it so both close and further neighboring observations matter when attempting to determine the value at the interpolation point, which would explain why both $\mathfrak{I}_{\text{Mean}}$ and $\mathfrak{I}_{\text{Median}}$ outperform $\mathfrak{I}_{\text{Nearest}}$.

Now about $\mathfrak{I}_{\text{\CCNDDNISD{}}}$, it simply outclasses the other interpolation functions, regardless of the data nature~$\Upsilon$ and number of nearest neighbors to consider~$k$ (except for the degenerate case where $k = 1$ where all these interpolation functions degenerate into nearest neighbor interpolation in which case they perform identically). This goes to show the utility of $\mathfrak{I}_{\text{\CCNDDNISD{}}}$'s realistic proximal weighting based on orthodromic distances and its \CCNDDlong{}. These allow $\mathfrak{I}_{\text{\CCNDDNISD{}}}$ to assign higher importance to observations very near the interpolation point while tuning down the importance of further (and potentially less relevant) neighbors, thus minimizing the amount of noise affecting the estimated values, which is where $\mathfrak{I}_{\text{\CCNDDNISD{}}}$'s robustness to noise originates from.

In summary, $\mathfrak{I}_{\text{\CCNDDNISD{}}}$ significantly outperforms the other interpolation functions in all experimental configurations (except for the degenerate $k = 1$ case where they are all expected to perform the same). Lastly, for anyone interested,~\cite{Trempe2019worldwide} discusses these results more extensively.

\CCMultiFigure{fig:sknni_res_rw}{%
	\CCMultiFigureSub{fig:interp_rw_temperature}{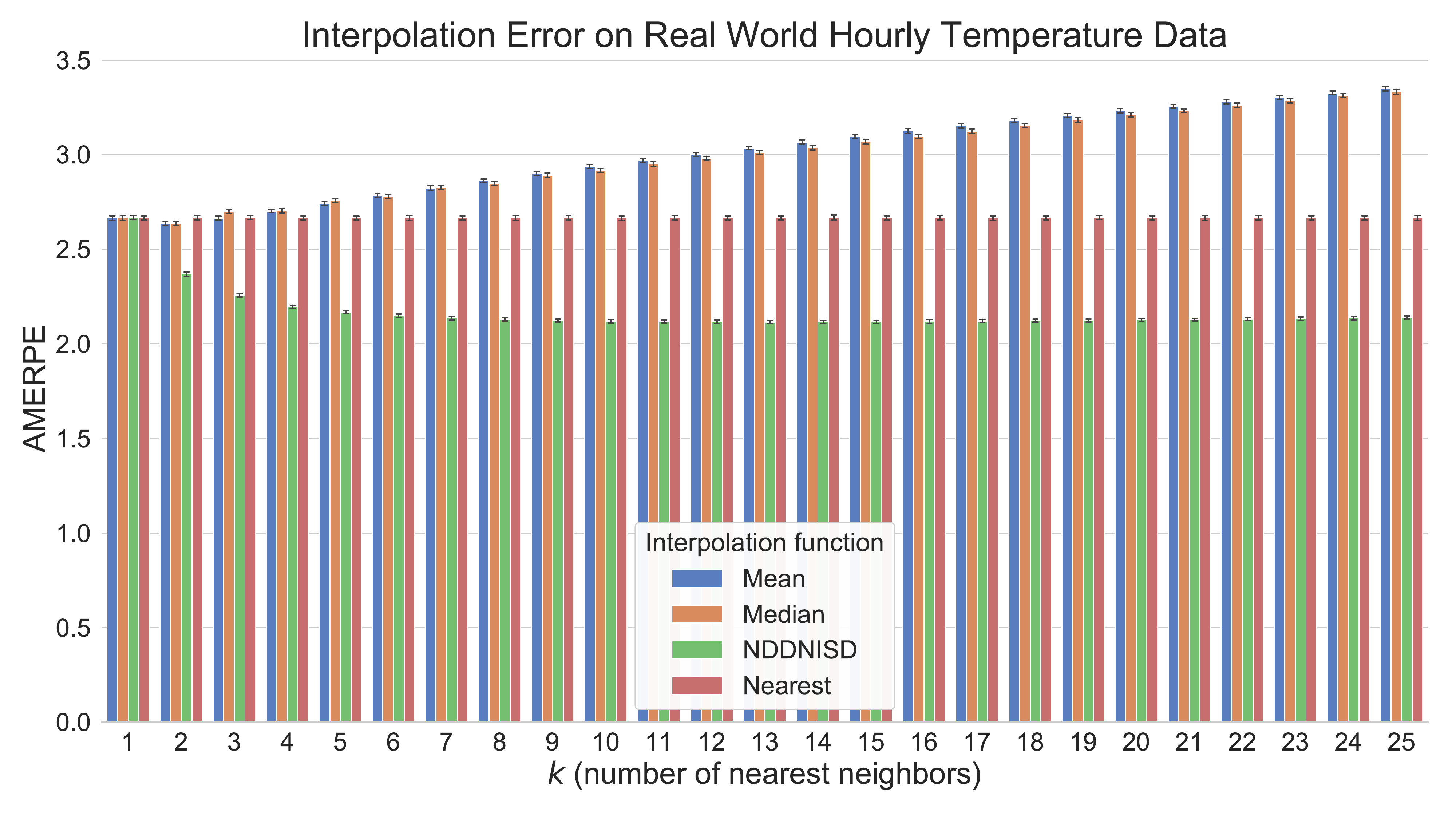}
		{Real World Temperature Data}{width=0.45\linewidth}
	\hfill
	\CCMultiFigureSub{fig:interp_rw_dew_point}{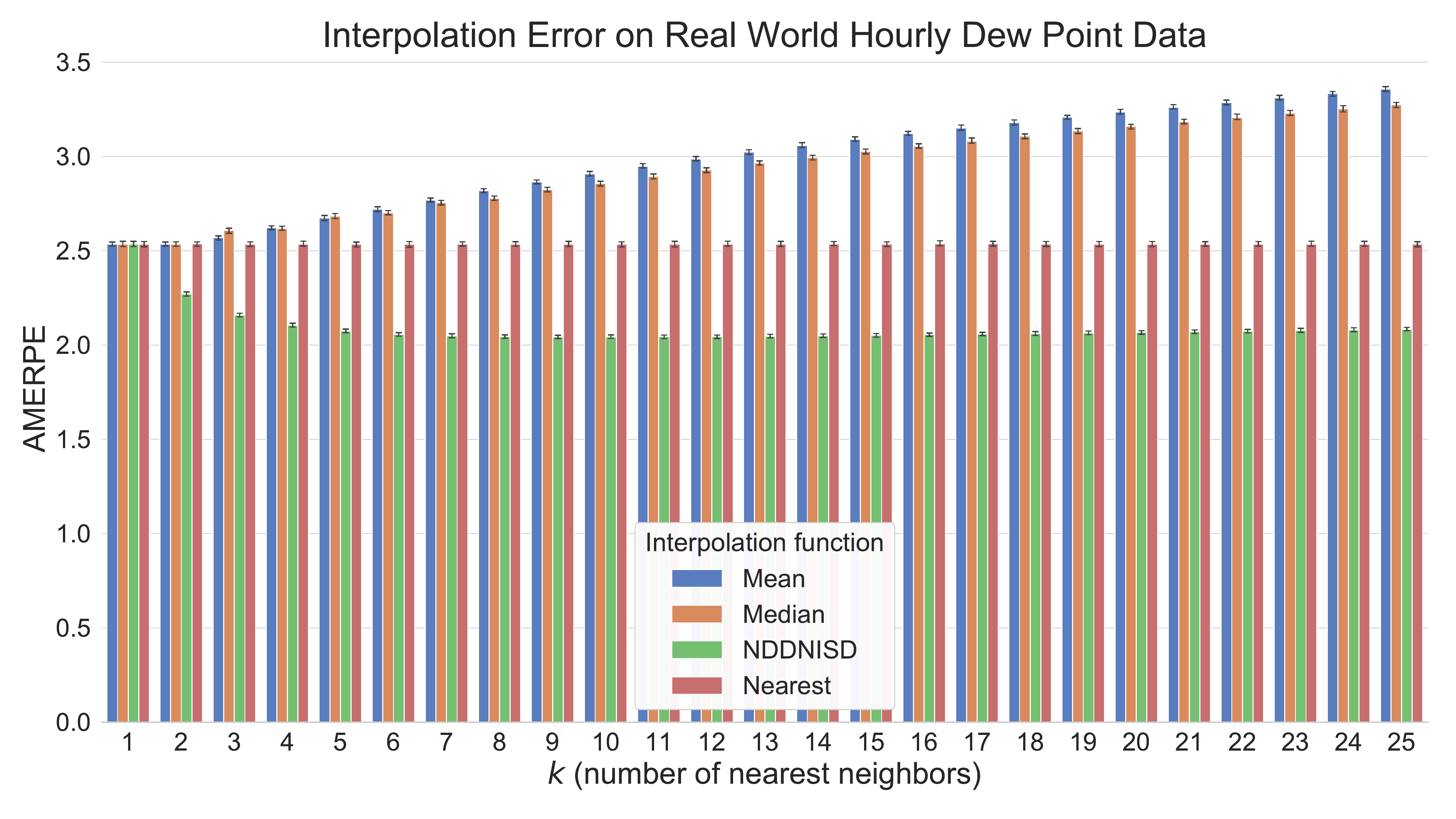}
		{Real World Dew Point Data}{width=0.45\linewidth}
	\\
	\CCMultiFigureSub{fig:interp_rw_pressure}{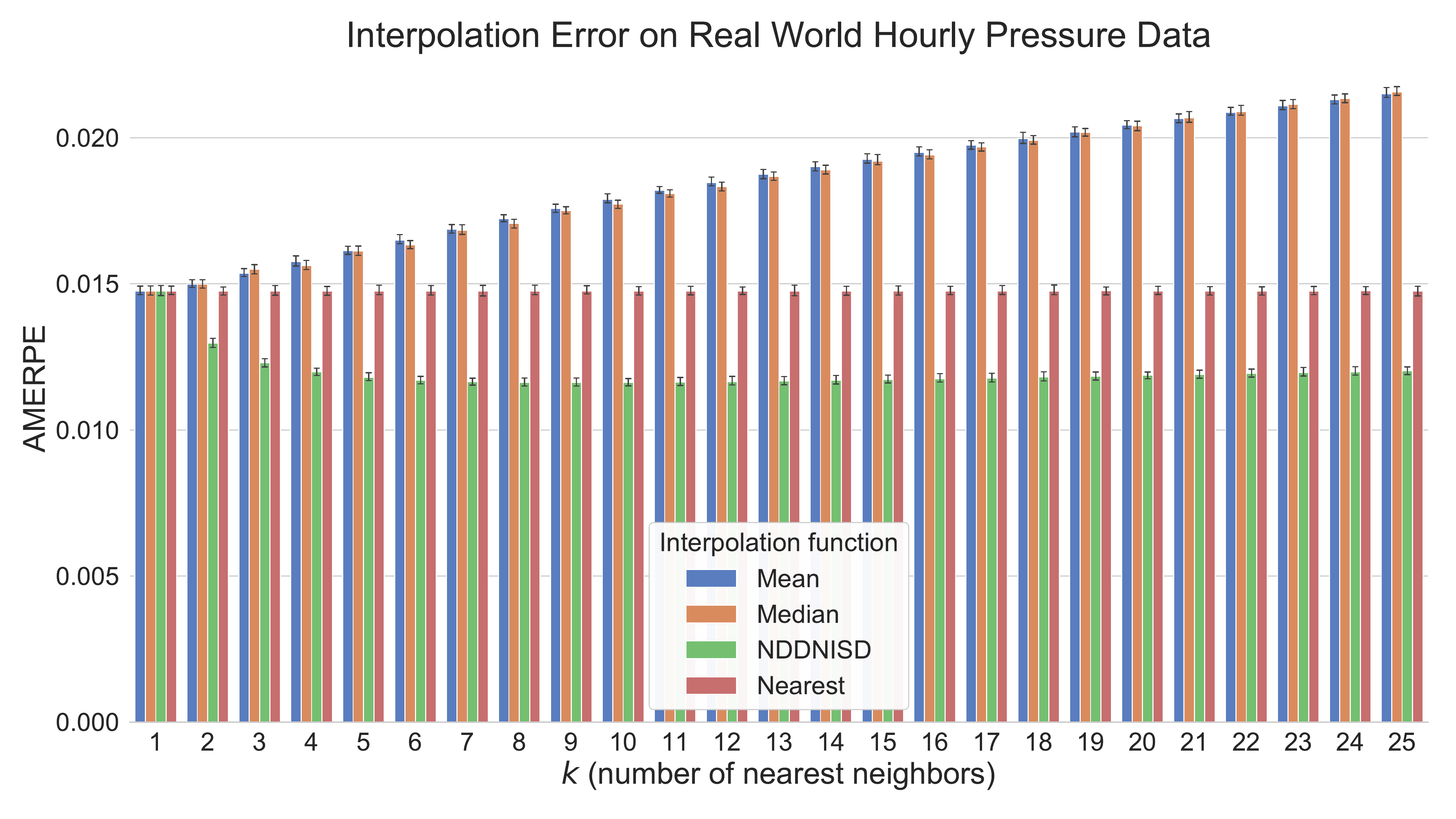}
		{Real World Pressure Data}{width=0.45\linewidth}
	\hfill
	\CCMultiFigureSub{fig:interp_rw_wind_speed}{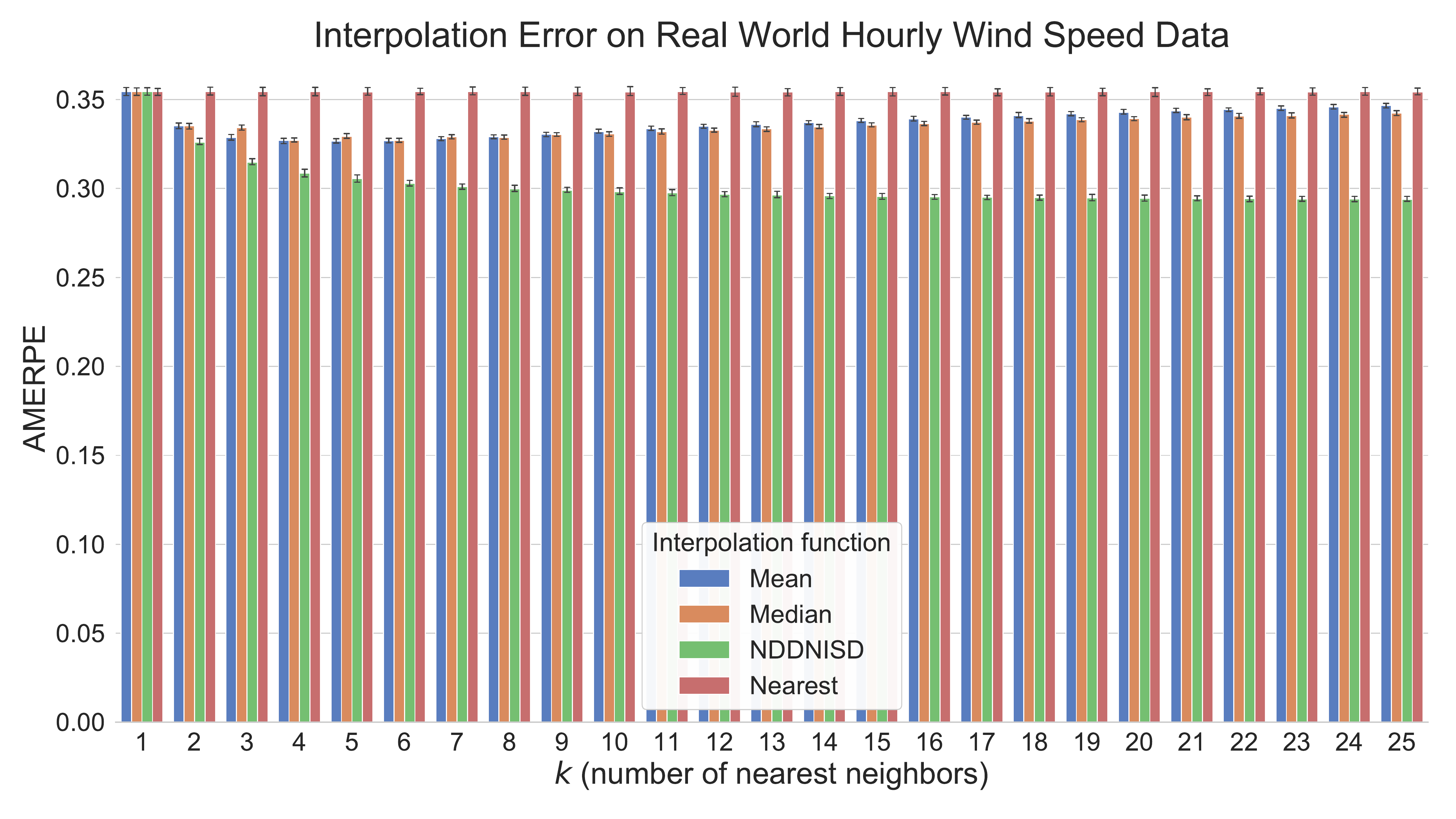}
		{Real World Wind Speed Data}{width=0.45\linewidth}}
{Interpolation Error on Real World Data}
{The figure shows the interpolation error of various interpolation functions on real world hourly weather data. For each weather parameter, \num{100} experiments were run, each for a different hourly observation set (sampled from a set of observation sets), and AMERPE was calculated for various interpolation functions for \num{25} values of $k$ (number of nearest neighbors) using \num{1000} sampled observations and about \num{3000} validation observations (the rest) that were held out for evaluation. The error bars represent the \SI{95}{\percent} confidence interval for the bootstrap mean (over \num{100} bootstrap samples) of the evaluated quantities.}

\section{Implementation}\label{sec:implementation}
As stated earlier, the implementation of \CCSkNNI{} (along with $\mathfrak{I}_{\text{\CCNDDNISD{}}}$) is open source and freely available online at \CCSknniRepositoryUrl{}. The library is implemented purely in Python with NumPy and SciPy as its sole dependencies. Furthermore, for the ones who only want to use \CCSkNNI{} as is, it can simply be added to any Python project by installing it through \texttt{pip} (Package Installer for Python) like so: \texttt{pip install sknni}, since \CCSkNNI{} is available on PyPI (Python Package Index). The ones wanting to adapt and modify \CCSkNNI{} for specific applications are of course encouraged to fork the project's repository and do so. To conclude this section, the code sample presented in \CCReflst{lst:sknni_usage_example} shows how simple \CCSkNNI{} actually is to use.
\begin{lstlisting}[
language=Python,
label=lst:sknni_usage_example,
caption=SkNNI Usage Example,
captionpos=t,
belowcaptionskip=0.5em,
float=!t,
floatplacement=!t,
upquote=true,
tabsize=4,
basicstyle=\selectfont\scriptsize\ttfamily,
numbers=left, 
numberstyle=\scriptsize, 
numbersep=1em, 
frame=single,  
xleftmargin=2.5em,
xrightmargin=5em,
framexleftmargin=2em]
import numpy as np

from sknni import SkNNI

if __name__ == '__main__':
    observations = np.array([[30, 120, 20],
                             [30, -120, 10],
                             [-30, -120, 20],
                             [-30, 120, 0]])
    interpolator = SkNNI(observations)
    interp_coords = np.array([[30, 0],
                              [0, -120],
                              [0, 0],
                              [0, 120],
                              [-30, 0]])
    interpolation = interpolator(interp_coords)
    print(interpolation)

# Output:
# [[  30.          0.          9.312546]
#  [   0.       -120.         14.684806]
#  [   0.          0.         12.5     ]
#  [   0.        120.         10.315192]
#  [ -30.          0.         16.464548]]
\end{lstlisting}

\section{Conclusion}\label{sec:conclusion}
Finally, this last section concludes this work by first presenting a summary of it, followed by its main contributions, and ends on considerations pertaining to future work.

\subsection{Summary}
Succinctly, to tackle the problem of sparse spherical interpolation, this work presents the \CCSkNNI{} geospatial interpolation algorithm and its \CCNDDNISD{} interpolation function. This work then evaluates \CCNDDNISD{} and compares it against commonly used interpolation functions. The experimental results show \CCSkNNI{}'s \CCNDDNISD{} significantly outperforms the other interpolation functions due to its spatial awareness, robustness to noise in observation values, and proximal weight renormalization based on neighborhood spatial distribution biases.

\subsection{Contributions}
This work's main contributions are as presented herein. The first major contribution of this work is \CCSkNNI{}, a spherical interpolation algorithm very well suited to work with sparse and irregular geospatial data, as is often encountered in real world scenarios. The second principal contribution of this work is \CCNDDNISD{}, an interpolation function for \CCSkNNI{} which performs well on both synthetic and various natures of real world data, and which shines due to its spatial (proximity and distribution) awareness. The third and last main contribution of this work is the open source implementation of \CCSkNNI{} and its \CCNDDNISD{} interpolation function, which shines due to achieving the initially sought ease of use (as shown in \CCReflst{lst:sknni_usage_example}), accuracy and flexibility.

\subsection{Future Work}
Ultimately, since this work aims to contribute to the advancement of sparse spherical data processing, anyone is welcome to integrate \CCSkNNI{} into their geospatial data processing pipelines and create their own adapted \CCSkNNI{} interpolation functions if needed. As such, it would be interesting to see how these new specific interpolation functions fair when compared to \CCNDDNISD{}. Another interesting kind of future work would be about creating a planar version of \CCSkNNI{}, which could currently be approximated by only using a small spatial region of a sphere with a huge radius. Lastly, exploring adaptations of \CCSkNNI{} to support categorical data, e.g. weather condition, is also a research avenue worth investigating due to its potential and practical applications.


%

%

\ifCLASSOPTIONcompsoc
  \section*{Acknowledgments}
\else
  \section*{Acknowledgment}
\fi
The author would like to thank Polytechnique Montréal, with a special thank to Michel Gagnon; and \CCPelmorexCorp{}, with a special thank to Edwin Vargas, for their support and collaboration in making the undertaking and achievement of this work possible.

\balance

\ifCLASSOPTIONcaptionsoff
  \newpage
\fi



\bibliographystyle{IEEEtran}
\bibliography{IEEEabrv,ms}
\end{document}